\documentclass[fleqn,usenatbib]{mnras}

\usepackage[T1]{fontenc}
\usepackage{ae,aecompl}
\usepackage[utf8]{inputenc}

\usepackage{graphicx}	
\usepackage{amsmath}	
\usepackage{amssymb}	

\newcommand{\re}{\hbox{${\rm R}_{\rm e}$}}

\newcommand{\mbh}{$M_\bullet$}
\newcommand{\msun}{\hbox{M$_{\odot}$}}

\title[Black hole heating]{Black hole feedback and the evolution of massive early-type galaxies}

\author[I. Mart\'in-Navarro]{
Ignacio Mart\'in-Navarro$^{1,2}$\thanks{E-mail: imartinn@ucsc.edu}, Joseph N. Burchett$^{1}$, \&  Mar Mezcua$^{3,4}$ 
\\
$^{1}$University of California Santa Cruz, 1156 High Street, Santa Cruz, CA 95064, USA\\
$^{2}$Max-Planck Institut f\"ur Astronomie, Konigstuhl 17, D-69117 Heidelberg, Germany\\
$^{3}$Institute of Space Sciences (ICE, CSCIC), Campus UAB, Carrer de Can Magrans, 08193, Barcelona, Spain\\
$^{4}$Institut d'Estudis Espacials de Catalunya (IEEC), C/ Gran Capit\`{a}, 08034 Barcelona, Spain\\
}

\date{Accepted XXX. Received YYY; in original form ZZZ}

\pubyear{2018}

\begin{document}
\label{firstpage}
\pagerange{\pageref{firstpage}--\pageref{lastpage}}
\maketitle

\begin{abstract}

Observationally, constraining the baryonic cycle within massive galaxies has proven to be quite difficult. In particular, the role of black hole feedback in regulating star formation, a key process in our theoretical understanding of galaxy formation, remains highly debated. We present here observational evidence showing that, at fixed stellar velocity dispersion, the temperature of the hot gas is higher for those galaxies hosting more massive black holes in their centers. Analyzed in the context of well-established scaling relations, particularly the mass--size plane, the relation between the mass of the black hole and the temperature of the hot gas around massive galaxies provides further observational support to the idea that baryonic processes within massive galaxies are regulated by the combined effects of the galaxy halo virial temperature and black hole feedback, in agreement with the expectations from the EAGLE cosmological numerical simulation.

\end{abstract}

\begin{keywords}
galaxies: formation -- galaxies: evolution  -- galaxies: abundances -- galaxies: stellar content
\end{keywords}


\section{Introduction}

Early-type galaxies (ETGs) dominate the massive end of the galaxy stellar mass function \citep[e.g.][]{Kelvin14}. Their relatively simple light profiles \citep[e.g.][]{Kormendy12} and the fact that their stellar populations can be accurately approximated by single stellar population (SSP) models \citep[e.g.][]{vazdekis96} have made ETGs observational benchmarks to study the formation and evolution of galaxies, both locally and at higher redshifts \citep[e.g.][]{Kriek16,MN18}.

ETGs have historically been thought to result from intense, quasi-dissipationless star formation events in the early Universe \citep[e.g.][]{Larson74}. This monolithic-like view of the formation process of ETGs is observationally supported by the stellar population properties of these objects. ETGs, and in particular their stellar populations, follow tight scaling relations with galaxy mass and/or velocity dispersion.  With increasing galaxy mass, ETGs are older, more metal-rich, and exhibit higher fractions of $\alpha$ to Fe-peak elements \citep[e.g.][]{thomas05,Johansson12}. Moreover,  there is now growing observational evidence suggesting that the stellar initial mass functions in ETGs also change systematically with galaxy mass \citep[e.g.][]{flb13}. This tight coupling between galaxy mass and stellar population properties seems to suggest an {\it in situ} formation of the central regions of massive ETGs, since the observed chemical properties of these objects result from star formation processes that are only sustainable in the center of massive dark matter halos \citep[e.g.][]{Matteucci94}.

This more traditional view of massive ETG formation is, however, at odds with a na\"ive interpretation of the favored $\Lambda$-Cold Dark Matter ($\Lambda$-CDM) cosmology. While the stellar population properties of ETGs show that more massive galaxies formed faster and earlier, the $\Lambda$-CDM scenario predicts a {\it bottom-up} structure growth where massive halos form later out of the accretion of smaller satellites \citep[e.g.][]{Blumenthal84}. This hierarchical formation process is not only supported by theoretical arguments, but also by observations suggesting that massive galaxies have grown in size over cosmic time \citep[e.g.][]{Daddi05,Trujillo06}.

Differences between the two formation scenarios, {\it top-down} vs {\it bottom-up}, are ultimately related to differences in the baryonic cycle. Under the {\it top-down} interpretation, stellar ejecta are recycled into new stars with an efficiency roughly determined by the balance between local escape velocity and the speed of the stellar winds  \citep[e.g.][]{Franx90}. More massive ETGs would have a deeper potential well, which would lead to a more efficient retention of chemically-enriched gas available to form new generations of stars. This would naturally produce massive and metal-rich ETGs and a relation between escape velocity and the chemical properties of massive ETGs \citep[e.g.][]{Davies93,Scott09}. However, from the point of view of current $\Lambda$-CDM numerical simulations, massive ETGs live within dark matter halos more massive than $\sim 10^{12}$ \msun, where the baryonic cycle is supposed to be dominated by the  effects of active galactic nuclei (AGN) feedback \citep[e.g.][]{Silk12}. In this case, it is expected that the chemical enrichment is regulated by the thermodynamics of stellar ejecta within the hot gaseous halos around massive galaxies. If these gaseous halos are hot enough, either via shock heating \citep[e.g.][]{Birnboim03} or due to the energy released by AGN activity \citep[e.g.][]{Sijacki15}, low-entropy stellar ejecta may be trapped by the high entropy surroundings,  eventually being recycled into new, metal-rich stars \citep{Bower17}. This hot corona would also prevent long-lasting star formation events fueled by accreting fresh cold gas, although the existence of a hot corona might not be enough to fully suppress star formation \citep[e.g.][]{Su19}.

It has been recently shown that star formation histories \citep{MN18b,MN18c}, chemical enrichment  \citep{MN16}, and present-day star formation rates \citep{Terrazas16,Terrazas17} within massive galaxies depend, at fixed galaxy mass or velocity dispersion, on the masses of the central super-massive black holes. These observational results suggest that, as expected from state-of-the-art cosmological simulations, the effect of black hole feedback plays a critical role in regulating star formation as it is thought to contribute to the heating of the gaseous envelope within massive halos. However, observationally, the link between black hole activity and the heating of the gaseous corona remains a matter of debate \citep{Bogdan15,Lakhchaura19,Gaspari19}. Does the temperature of the gas surrounding massive galaxies correlate with the mass of the central super-massive black holes? Or, in the context of the {\it top-down} vs {\it bottom-up} scenarios, does stellar ejecta {\it fall} towards the centers of massive galaxies because its velocity is lower than the escape velocity or does it {\it sink} because its entropy is lower than that of the surrounding gaseous halo?

In this work, we explore the connection between super-massive black holes, stellar population properties, and hot gas temperature in the context of massive galaxy evolution. At fixed stellar velocity dispersion we find a clear correlation between X-ray temperature and black hole mass, as presented in \S~\ref{sec:temp}. In \S~\ref{sec:mss}, we propose an alternative method to study the  black hole--galaxy connection when direct black hole mass measurements are not available. The implications of these findings are discussed in \S~\ref{sec:disc}, and the conclusions are summarized in \S~\ref{sec:fin}.

\section{The T$_\mathrm{gas}$ -- $\sigma$ relation} \label{sec:temp}

In order to study the interplay between gas temperature and black hole activity, we followed the same approach as in \citet{MN16}. In short, the basic idea consists of using the relation between black hole mass and galaxy velocity dispersion (M$_\bullet$-$\sigma$) as a metric for the amount of feedback injected into a given galaxy. Assuming that the energy radiated by a growing black hole scales with its mass, and that a fraction of that energy gets coupled to the gas in the galaxy \citep[e.g.][]{Sijacki07,Booth:2009aa}, more energy would have been injected via AGN activity in those objects hosting more massive black holes. In practice this means that the distance of a galaxy from the average M$_\bullet$-$\sigma$ relation ($\Delta$M$_\bullet$) directly correlates with the (relative) amount of black hole feedback experienced during its evolution, independently from the initial mass of the black hole seeds \citep{Mezcua17}. Therefore, by comparing the properties of galaxies above and below the M$_\bullet$-$\sigma$ relation, so-called over-massive and under-massive black hole galaxies, respectively, one can empirically assess the effect of black hole feedback on the baryonic cycle of galaxies.  

\subsection{Stellar masses, black holes, and X-ray temperatures}
We base our analysis on the sample of black hole masses presented in \citet{vdb16}, consisting of 230 galaxies where black hole masses have been dynamically measured. All the galaxies in the sample are at $z \approx 0$, and they cover a stellar mass range from $\sim 10^{10}$ to $\sim 10^{12}$ \msun. Most of these objects are ETGs, particularly on the massive end \citep{vdb16}. In addition to black hole masses, the catalog presented in \citet{vdb16} also provides stellar velocity dispersion measurements at a fixed aperture of one effective radius \re. Stellar masses are also given thanks to an homogeneous analysis of $K$-band imaging from the Two Micron All Sky Survey \citep{Skrutskie06}. Throughout the rest of this work, all the discussion is based on the stellar velocity dispersions and masses listed in \citet{vdb16}.

We crossmatched this sample with two catalogs of X-ray temperature measurements, \citet{Kim15} and \citet{Goulding16}. Both works measured the temperature of the hot gaseous component within $\sim$1\re \ by fitting the X-ray spectrum in the $\sim$ 0.3 to 5 keV energy range. The \citet{Kim15} sample focused on the ATLAS3D project \citep{Cappellari11}, a volume-limited survey of nearby ETGs. Complementarily, \citet{Goulding16} presented an X-ray study of the MASSIVE survey \citep{Ma14}, which complements the ATLAS3D project by including galaxies with stellar masses above 10$^{11}$ \msun. Thus, these measurements probe the hot gaseous corona within the stellar component rather than in the galaxy halos. Our final sample consists of 50 galaxies with direct black hole mass, stellar velocity dispersion, and X-ray gas temperature measurements. Out of these 50 galaxies, 37 belong to the ATLAS3D project, while the remaining 17 were drawn from the MASSIVE sample. Moreover, only 12 objects belong to the Virgo cluster, while the others reside in less dense environments. Note that, because of the selection functions of both surveys, ATLAS3D and MASSIVE probe different mass regimes. While ATLAS3D contains numerous relatively low-mass, fast-rotators, MASSIVE is heavily biased towards the massive, slow-rotator ETG population. Therefore, the two samples probe not only different sets of galaxies but also galaxies that have likely undergone different evolutionary processes.

\subsection{Dependence of T$_\mathrm{gas}$ on black hole mass}

\begin{figure}
\begin{center}
\includegraphics[width=8.cm]{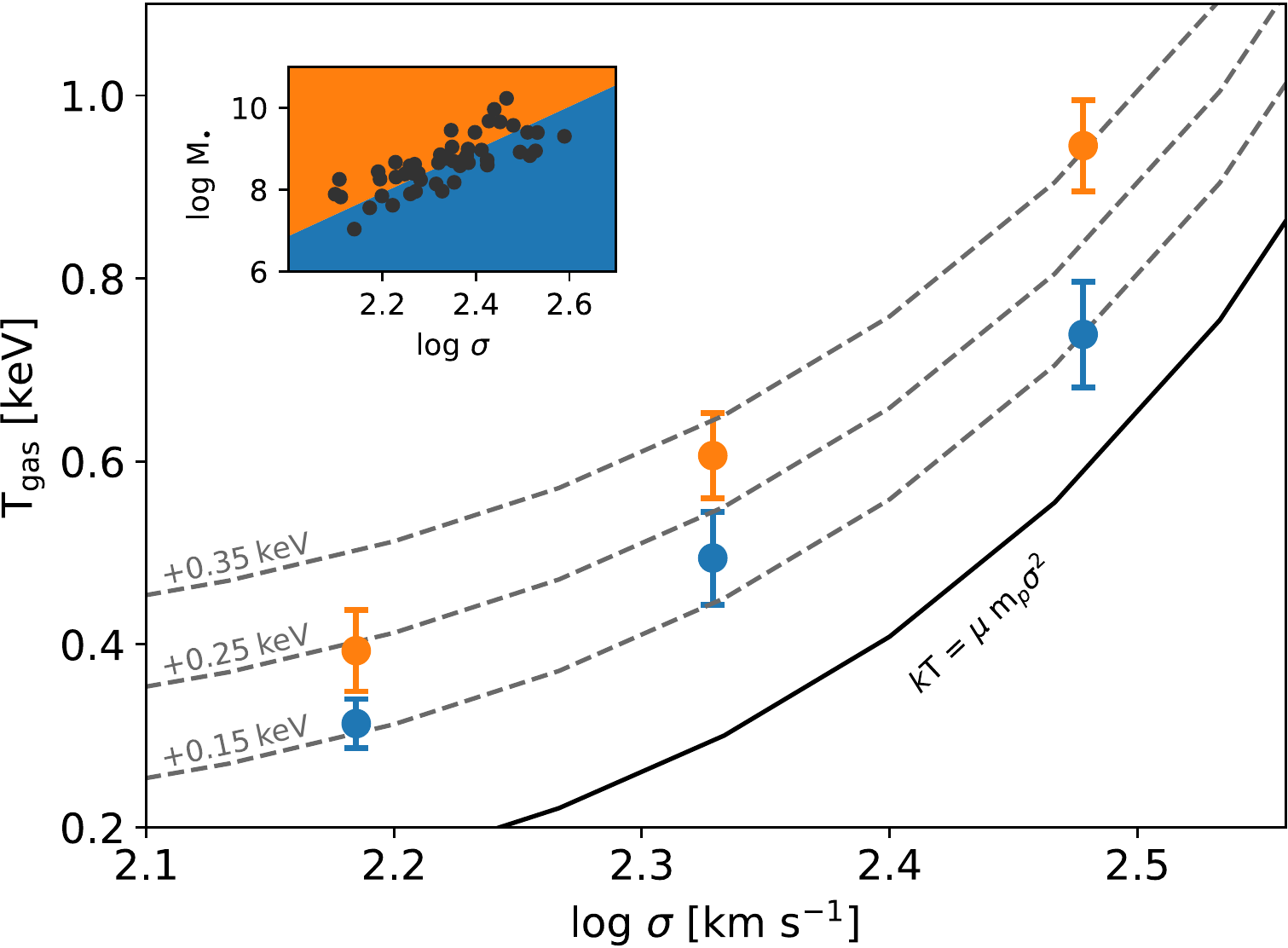}
\end{center}
\caption{The T$_\mathrm{gas}$ -- $\sigma$ relation for our sample of over-massive (orange) and under-massive (blue) black hole galaxies. At fixed velocity dispersion, galaxies above the average M$_\bullet$-$\sigma$ have X-ray emitting gas that is hotter than those in under-massive black hole galaxies. The solid black line marks the limit where gas and stars have the same kinetic energy, while dashed lines indicate +0.15, +0.25, and +0.35 keV increments in the temperature of the gas with respect to the idealized kinetic energy implied by the stars. The temperature difference between over- and under-massive black hole galaxies clearly scales with the mass of the black hole, suggesting a connection between black hole growth and coronal gas temperature at fixed velocity dispersion. The inset shows the M$_\bullet$-$\sigma$ relation of our sample of 50 objects, with the best-fitting relation indicated with a solid line.}
\label{fig:tgas}
\end{figure}

In Figure~\ref{fig:tgas} we show the observed dependence of the coronal gas temperature as a function of galaxy velocity dispersion for over-massive and under-massive black hole galaxies. The overall trend, as previously reported, follows a relation of roughly T$_\mathrm{gas} \sim \sigma^{1.5}$ \citep[see, e.g.,][]{Davis96}. In addition, the observed T$_\mathrm{gas}$ values for both over-massive and under-massive black hole galaxies are higher than expected from virial arguments, indicating that additional mechanisms contribute to heating the gas \citep{Davis96,Goulding16}.

However, the most striking feature shown by Figure~\ref{fig:tgas} is the clear separation between over-massive and under-massive black hole galaxies. Galaxies hosting more massive black holes exhibit systematically higher T$_\mathrm{gas}$. Moreover, the temperature difference between over- and under-massive black hole galaxies is a strong function of galaxy velocity dispersion and therefore of black hole mass. For galaxies with $\log \sigma \sim 2.2$ km s$^{-1}$ the temperature difference is $\sim 0.1$ keV. In contrast, for the most massive galaxies in our sample, with $\log \sigma \sim 2.5$, over-massive black hole galaxies have coronal gas $\sim 0.3$ keV hotter than under-massive black hole galaxies.  A more quantitative analysis of Figure~\ref{fig:tgas} reveals that 

\begin{equation} \label{eq:heat}
     k \Delta \mathrm{T}_\mathrm{gas} \ \mathrm{[keV]} = 10^{-3.47 (\pm 0.35)} \cdot \Delta \mathrm{M}^{0.29 (\pm 0.03)}_\bullet \ \mathrm{[\msun]}
\end{equation}

\noindent where $\Delta \mathrm{T}_\mathrm{gas}$ and $\Delta \mathrm{M}_\bullet$ are the differences in gas temperature and in black hole mass between over-massive and under-massive black hole galaxies at fixed velocity dispersion, respectively, $k$ is the Boltzmann constant, and the quoted uncertainties in the coefficients correspond to the formal $1\sigma$ errors. Eq.~\ref{eq:heat} models the differential change in coronal gas temperature and black hole mass at a given stellar velocity dispersion, and suggests that heating due to black hole feedback scales (sub-linearly) with black hole mass. An alternative way to express the relation between T$_\mathrm{gas}$ and M$_\bullet$ for our sample of galaxies is given by the following best-fitting equation:

\begin{equation} \label{eq:direct}
     k \mathrm{T}_\mathrm{gas} \ \mathrm{[keV]} = 10^{-2.14 (\pm 0.22)} \cdot \mathrm{M}^{0.22 (\pm 0.02)}_\bullet \ \mathrm{[\msun]}
\end{equation}

\noindent where T$_\mathrm{gas}$ and M$_\bullet$ refer to coronal gas temperatures and black hole mass measurements for individual galaxies in our sample, respectively. As before,  $k$ is the Boltzmann constant and uncertainties in Eq.~\ref{eq:direct} are $1\sigma$ formal errors.

Fig.~\ref{fig:tgas} and its parametrization in Eqs.~\ref{eq:heat} and \ref{eq:direct} reveal a strong correlation between X-ray gas temperatures, galaxy velocity dispersions and black hole masses in our sample of low-$z$ massive galaxies. Moreover, the fact that X-ray temperatures more closely correlate with black hole masses rather than with the stellar velocity dispersions suggest that the black holes likely have an important role in controlling the internal temperatures of massive halos. It is worth emphasizing again that these X-ray temperature measurements are limited to the central ($r\sim$\re) regions of our galaxies and therefore not directly related to the circumgalactic medium (CGM). However, observations show that galaxies with hotter X-ray gas in their inner regions also exhibit hotter CGMs out to $r\sim10$\re \ \citep[e.g.][]{Fukazawa06,Diehl08}, supporting the idea that the thermodynamical properties of the outer CGM are, at least, partially coupled to those of the inner gaseous corona probed by Fig.~\ref{fig:tgas}.

\section{Beyond the M$_\bullet$-$\sigma$ relation} \label{sec:mss}

Studies coupling the baryonic properties of galaxies with the mass of the central super-massive black hole are providing compelling observational evidence for the role of black hole feedback in regulating the internal thermodynamics of massive dark matter halos, and therefore star formation. However, the analysis presented in Fig.~\ref{eq:heat} can only be done at $z\sim0$ as it faces an unavoidable limitation: direct black hole mass measurements are restricted to relatively nearby systems in the local Universe where the sphere of influence of the black hole can be resolved. This not only limits the number of objects feasibly studied but also might introduce biases in the analysis \citep[e.g.][]{Shankar19}. In order to provide a coherent and comprehensive observational picture, it is then necessary to investigate alternative metrics that do not require direct black hole mass measurements. 

\begin{figure}
     \begin{center}
     \includegraphics[width=8.7cm]{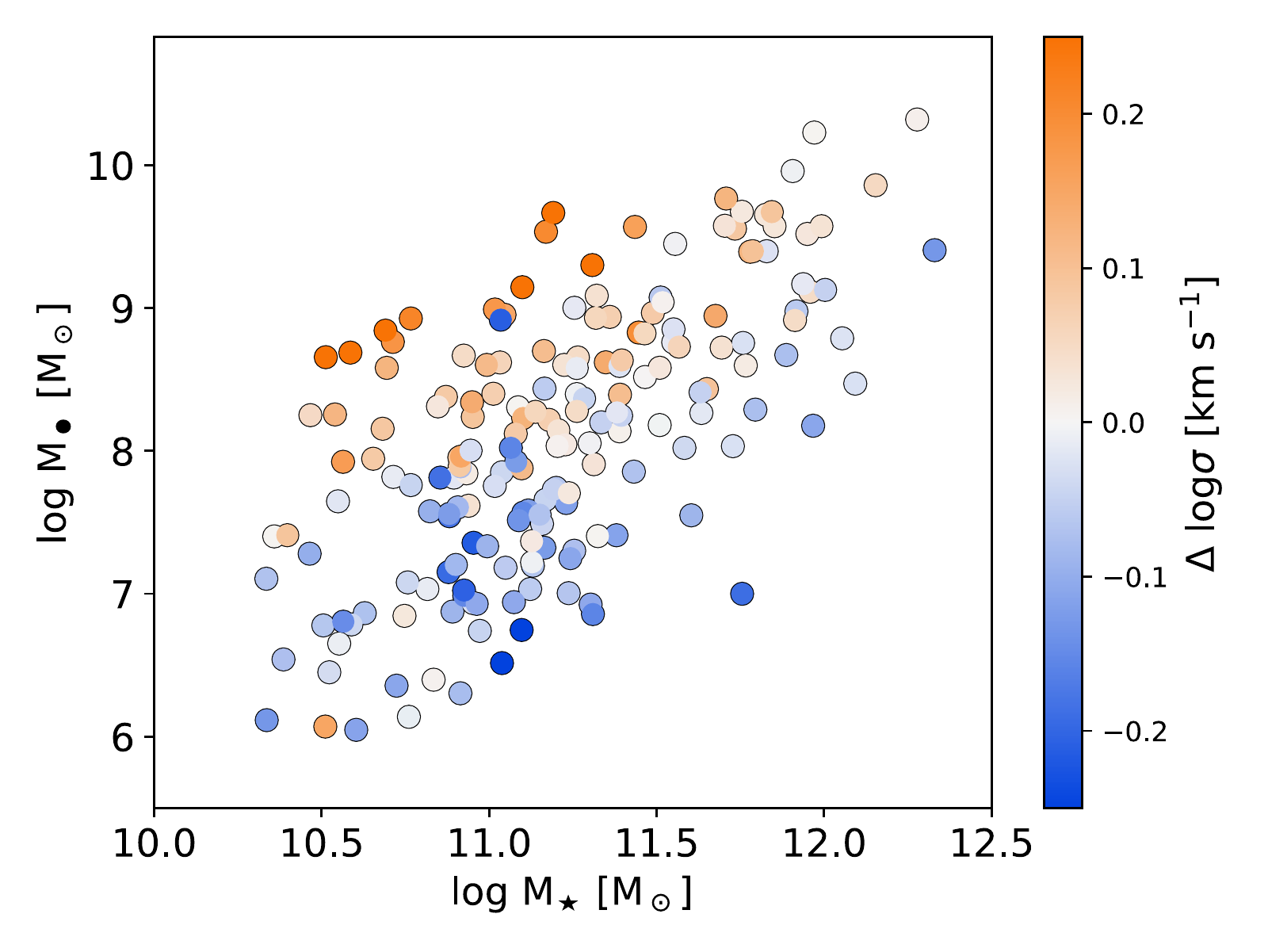}
     \end{center}
     \caption{The M$_\bullet$-$\sigma$-M$_\star$ relation for the \citet{vdb16} sample. Each symbol corresponds to a galaxy with a direct black hole mass measurement from the sample of \citet{vdb16}, color-coded as a function of its deviation with respect to the average $\sigma$-M$_\star$ relation. At fixed stellar mass, galaxies with more massive black holes in their centers also exhibit higher stellar velocity dispersion than the average.}
     \label{fig:scatter}
\end{figure}

Figure~\ref{fig:scatter} shows the M$_\bullet$-M$_\star$ relation based on the full sample of \citet{vdb16}. Each galaxy is color coded by $\Delta \log \sigma$, i.e., the difference between its measured velocity dispersion and the expected velocity dispersion given its stellar mass (determined by the best-fitting $\sigma$--M$_\star$ relation, see Appendix~\ref{app:1}). From Figure~\ref{fig:scatter} it becomes clear that, at fixed stellar mass, galaxies with more massive black holes also have higher velocity dispersions. This is ultimately due to the fact that black hole mass correlates more strongly with $\sigma$ than with stellar mass \citep[e.g.][]{Ferrarese00,Gebhardt00,Gultekin09,Beifiori12,vdb16}. In principle, this could be a powerful tool because it implies that it is not necessary to directly measure black hole masses in order to study the effect of black hole feedback on their host galaxies \citep[e.g.][]{Yu02,Marconi04,McLure04}. 

\begin{figure*}
     \begin{center}
     \includegraphics[width=7.cm]{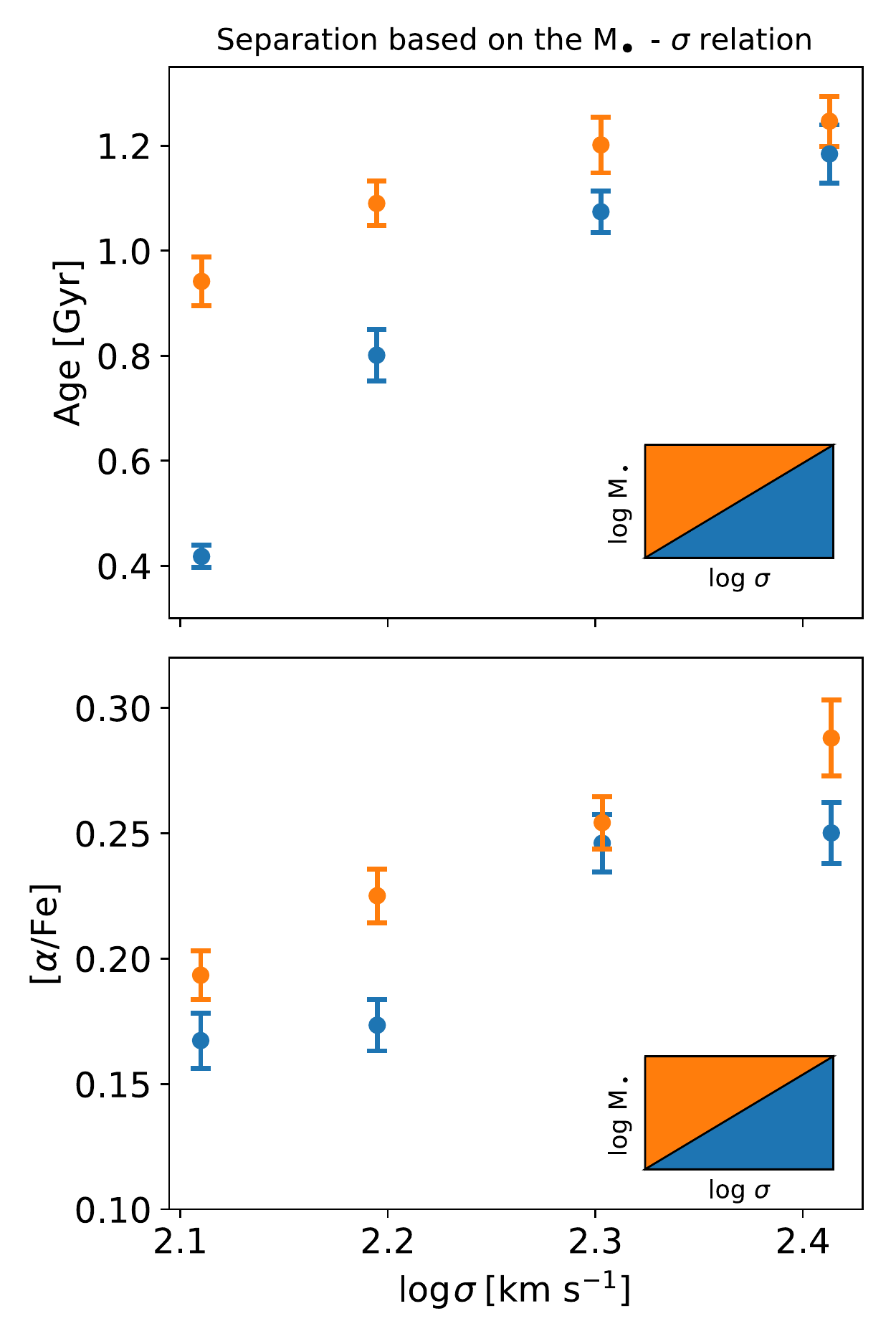}
     \includegraphics[width=7.cm]{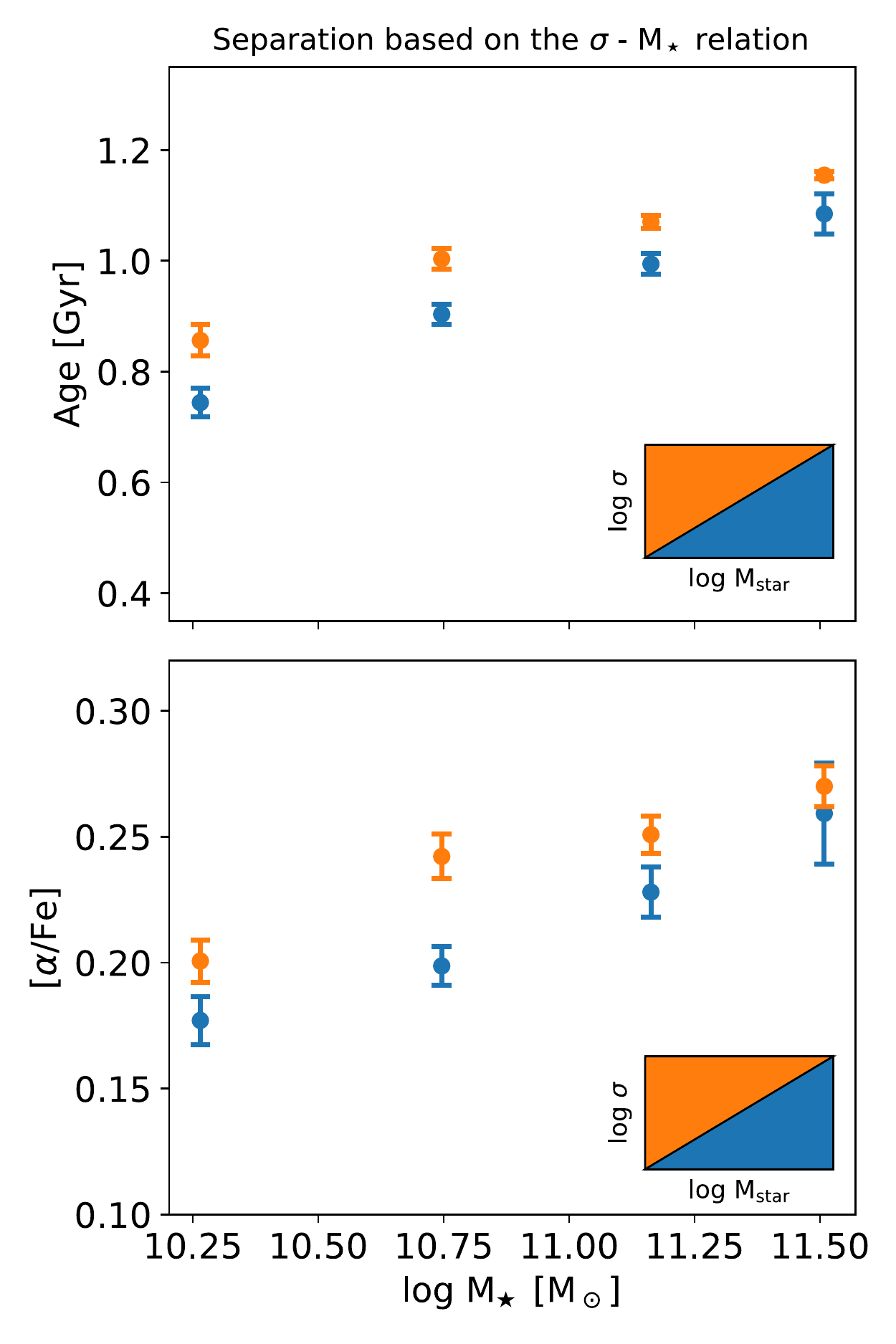}
     \end{center}
     \caption{On the left, adapted from \citet{MN16}, age-$\sigma$ (top) and [$\alpha$/Fe]-$\sigma$ (bottom) trends are shown for over-massive and under-massive black hole galaxies. The distinction between over- and under-massive was based on direct black hole mass measurements. On the right, equivalent trends from the ATLAS3D survey \citep{McDermid15}, where over-massive and under-massive black hole galaxies are those with high and low velocity dispersions, respectively. The fact that left and right panels show consistent trends further supports the idea of using stellar velocity dispersions in large and representative samples if direct black hole mass measurements are not available.}
     \label{fig:comp}
\end{figure*}

Before further applying $\sigma$ as a proxy for black hole mass, it is necessary to show that this approach leads to reasonable results. Figure~\ref{fig:comp} shows age and [$\alpha$/Fe] trends with galaxy velocity dispersion and stellar mass for over-massive and under-massive black hole galaxies based on two different approaches. In these plots, ages are SSP-equivalent measurements, while [$\alpha$/Fe] indicates the mass ratio of $\alpha$ elements to iron. Because the latter is mainly produced in Type Ia supernovae and $\alpha$ elements are synthesized in very massive, short-lived stars, [$\alpha$/Fe] effectively probes the duration of star formation in galaxies \citep[][although a certain IMF slope has to be assumed, \citealt{MN16b}]{thomas05}. High and low [$\alpha$/Fe] ratios indicate short and long star formation events, respectively.

The left panels in Figure~\ref{fig:comp} show the differences between over-massive and under-massive black hole galaxies as a function of velocity dispersion. The data points come from \citet{MN16} and the separation between over- and under-massive black hole galaxies is based on direct black hole mass measurements. On the right, we show the age and [$\alpha$/Fe] estimations from the ATLAS3D survey \citep{McDermid15} as a function of stellar mass. In this case, the distinction between over-massive and under-massive is  based on the M$_\star$-$\sigma$ relation. The similarities between the left and right panels are clear, showing that varying the black hole mass at fixed velocity dispersion leads to a change in the stellar population properties similar to that observed when varying the velocity dispersion at fixed stellar mass\footnote{Age differences in \citet{MN16} are larger than those found in \citet{McDermid15} because the former used line-strength indices, which are more biased towards young stellar populations. Note also that while ATLAS3D is limited to ETGs, the \citet{MN16} sample includes later types, in particular for lower-mass galaxies.}.

The use of proxies to estimate black hole masses in large samples of galaxies is obviously not a new idea, and has been extensively used to explore the co-evolution between galaxies and black holes \citep[e.g.][]{Benson07,Kelly12,Lang14,Bluck14,Woo15,Davor18}. Figure~\ref{fig:comp} further develops this idea by showing that, to first order, one can use $\sigma$ to probe galaxies with different black hole masses (at fixed stellar mass). However, some important caveats must be considered when following this approach. First, the intrinsic scatter in the M$_\bullet$-$\sigma$ relation is not zero \citep[e.g.][]{vdb16}. Therefore, a simple change in $\sigma$ at a given stellar mass does not fully account for the total change in black hole mass and its possible effect on the star formation histories \citep{MN18b}. Second, changes in velocity dispersion are also associated with changes in the internal structure of galaxies \citep[e.g.][]{Graham13,Sahu19} and even due to changes in orientation \citep[e.g.][]{Xiao11}. Therefore, discussing the effect of black hole feedback based on velocity dispersion measurements should be cautiously treated, controlling for any possible confounding variables. Finally, the mapping between black hole mass and $\sigma$ might change with galaxy mass \citep[e.g.][]{Davor18,Shankar19} and even with redshift. Thus, interpreting trends as a function of galaxy or halo mass might lead to biased results \citep{Bernardi07}.

\section{Discussion} \label{sec:disc}

There is now growing evidence suggesting that the baryonic properties of massive galaxies, namely their ages, star formation rates, and chemical composition, are tightly coupled to the masses of their central super-massive black holes \citep{MN16,MN18b,MN18c,Terrazas16,Terrazas17}. The observed relation between black holes and coronal gas temperature described in \S~\ref{sec:temp} provides further observational evidence for a scenario where the star formation in massive galaxies is regulated by the energetic input from the central black hole. In this scenario, the energy released by a black hole during its growth couples with the hot gaseous corona around massive galaxies, heating the gas above the expected virial temperature. This heating is proportional to the mass of the black hole and would naturally explain the observed properties of ETGs.

Numerical simulations suggest that a {\it ballistic} scenario, where stellar ejecta are recycled into new stars in a process regulated by the escape velocity of the galaxy, is not feasible for galaxies more massive than M$\gtrsim 10^{10.5}$ \msun, where the effect of AGN feedback is expected to become dominant \citep{Scannapieco15,Bower17,Nelson19}. Our results, summarized in Figure~\ref{fig:tgas}, favor a scenario where the baryonic cycle within massive galaxies is driven by a thermodynamic equilibrium between stellar ejecta and the surrounding hot corona \citep{Bower17}, whose temperature is controlled by both the mass (or virial temperature) of the dark matter halo and the energy input from the central super-massive black hole. This picture, combined with the realization that the M$_\bullet$-M$_\star$ plane can be approximated by the $\sigma$-M$_\star$ plane (although with important caveats as discussed in \S~\ref{sec:mss}), have direct implications on our understanding of ETGs.

\subsection{Re-thinking the mass--size plane}

\begin{figure}
     \begin{center}
     \includegraphics[width=8.7cm]{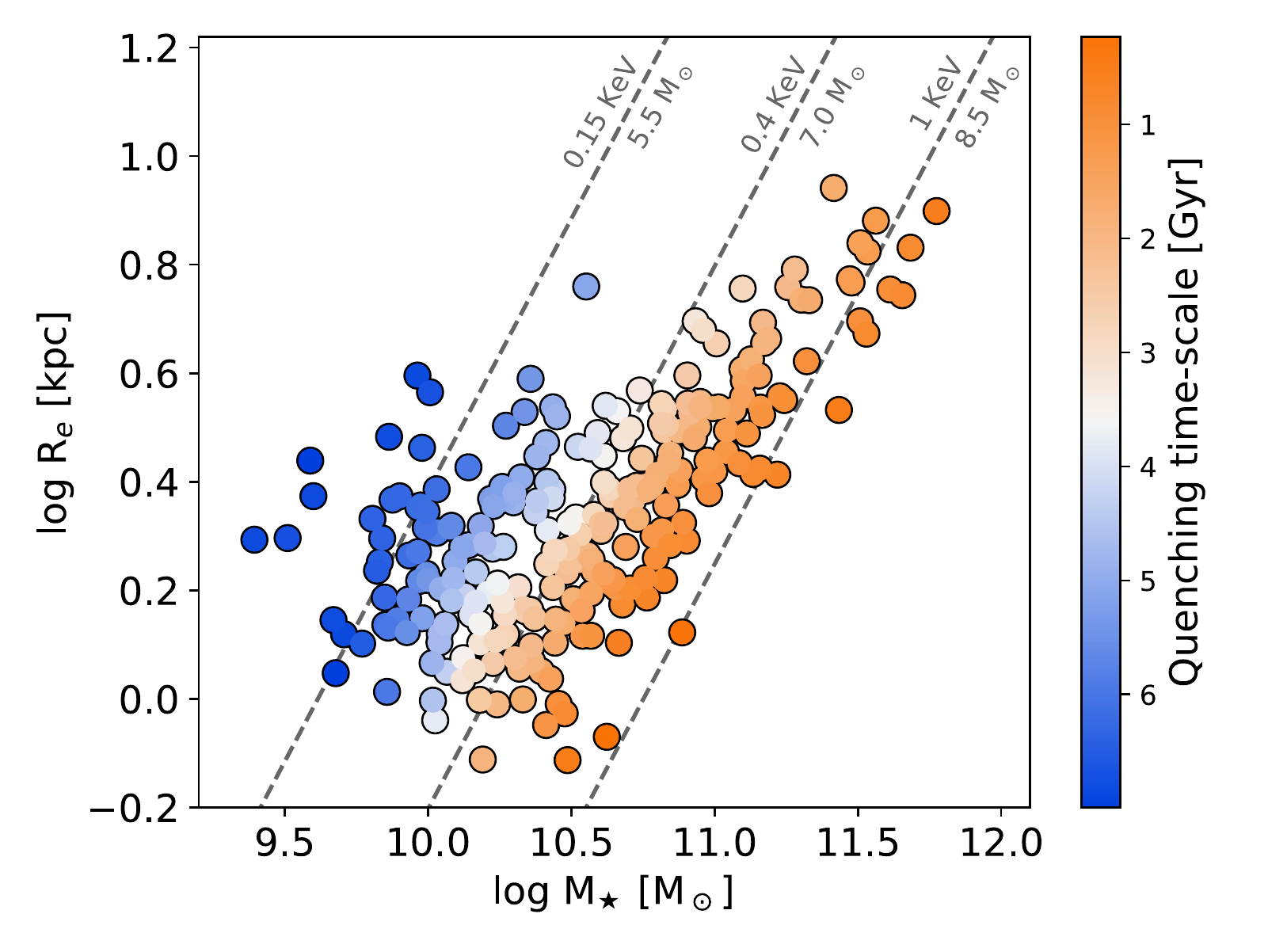}
     \end{center}
     \caption{Mass -- size plane of ETGs from the ATLAS3D survey \citep{Cappellari13b}, color-coded by quenching time-scale \citep{McDermid15} using the LOESS smoothing implementation described in \citet{Cappellari13}, as a proxy for the effect of black hole feedback on regulating star formation. Diagonal dashed lines indicate lines of constant velocity dispersion, which closely track the observed changes in galaxy properties across the mass -- size plane. Following Figure~\ref{fig:tgas} and Figure~\ref{fig:scatter}, lines of constant $\sigma$ can be approximated as lines of constant M$_\bullet$, and therefore of constant gas temperature (as labeled on each diagonal line). We hypothesize that the physical mechanism behind the observed stellar population trends in the mass -- size plane is actually a combination of potentially accreting gas being shock-heated to the virial temperature of the halo and the additional energy released from the central super-massive black hole. Note that these are all of the ETGs in the ATLAS3D sample, not our sub-sample shown in Figure~\ref{fig:tgas}.}
     \label{fig:msize}
\end{figure}

The mass--size plane, popularized by the ATLAS3D team \citep[e.g.][]{Cappellari13b}, can be understood as a projection of the size-luminosity-velocity dispersion space, i.e., of the fundamental plane of ETGs \citep[e.g.][]{Bender92}, and it is a powerful framework for understanding the formation and evolution of massive galaxies. One of the most striking features of the mass--size plane is the fact that the properties of galaxies smoothly transition following lines of constant velocity dispersion \citep[e.g.][]{Cappellari13,Cappellari16}. This is exemplified in Figure~\ref{fig:msize}, where the mass--size plane from the ATLAS3D survey is shown, color-coded by the quenching time-scales of galaxies \citep[defined as t$_{50}$ in][]{McDermid15}. Dashed lines of constant velocity dispersion are shown in grey, assuming that galaxies are virialized following $M_\star=5 R_e \sigma / G$ \citep{Cappellari13}, where G is the gravitational constant.

The combination of Figure~\ref{fig:tgas} and Figure~\ref{fig:scatter} allows us to propose that, in the mass--size plane, the stellar population properties of galaxies \citep[e.g. abundance pattern, age, and metallicity, see ][]{McDermid15} follow lines of constant $\sigma$, which, given the tight relation between $\sigma$, M$_\bullet$, and T$_\mathrm{gas}$ (Eq.~\ref{eq:direct}), can be also interpreted as lines of constant gas temperature (note the quantities labeled accordingly in Figure~\ref{fig:msize}). Moreover, Fig.~\ref{fig:tgas} suggests that this gas temperature is ultimately determined by the mass (or virial temperature) of the galaxy plus the additional energetic input from the central super-massive black hole. The apparent trends between stellar population properties and galaxy velocity dispersion \citep{McDermid15,Scott17,Li18} would be then just a consequence of the fact that $\sigma$ strongly depends on both halo and black hole mass, while the stellar mass of a galaxy is mostly just probing the mass of its dark matter halo. The role of black hole heating in regulating the properties of galaxies across the mass--size plane, and in particular the quenching efficiency, is further supported by the fact that the fraction and amount of cold gas decreases with increasing velocity dispersion \citep{Cappellari13,Davis19}.

\begin{figure*}
     \begin{center}
     \includegraphics[width=8.7cm]{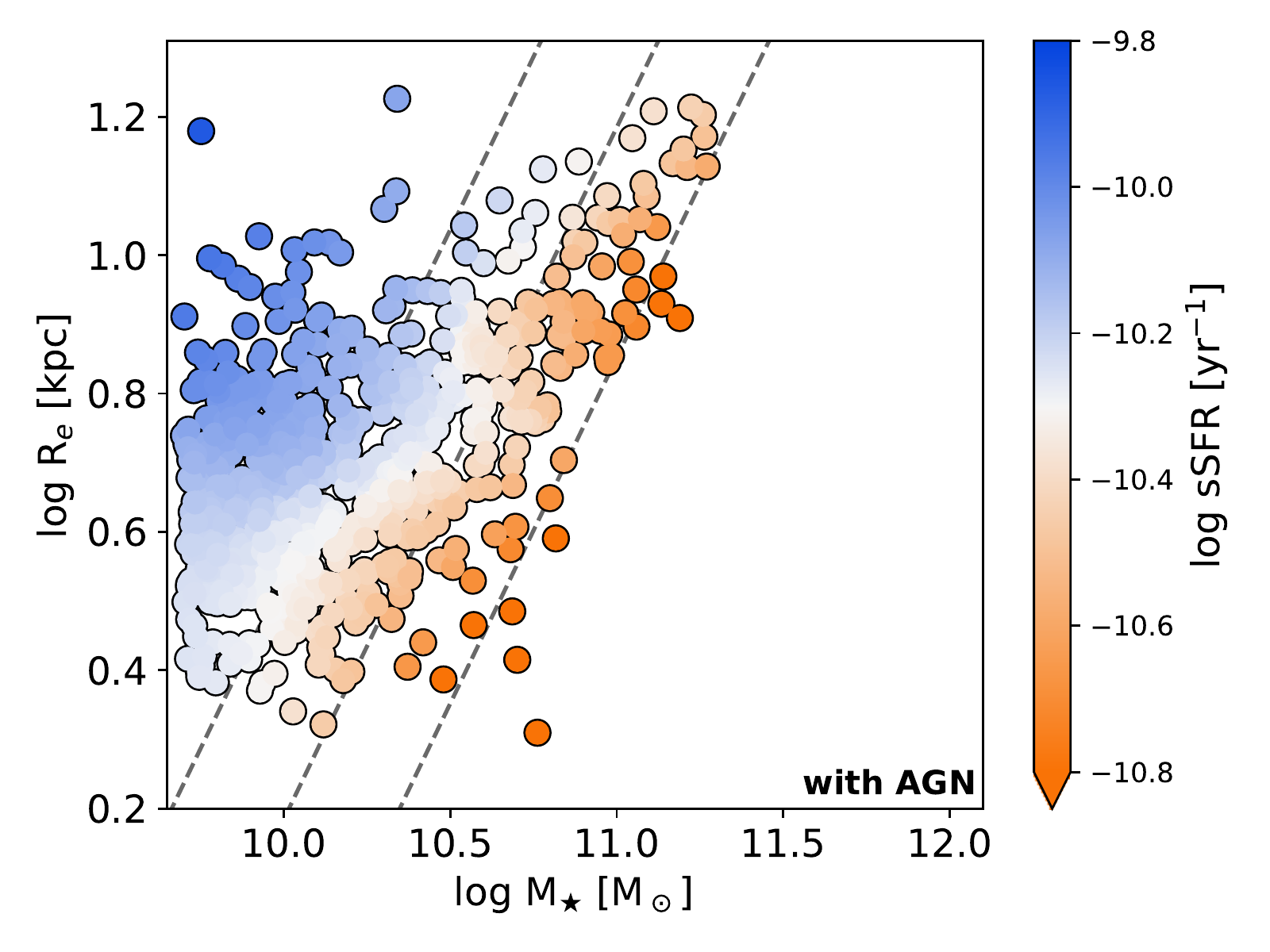}
     \includegraphics[width=8.7cm]{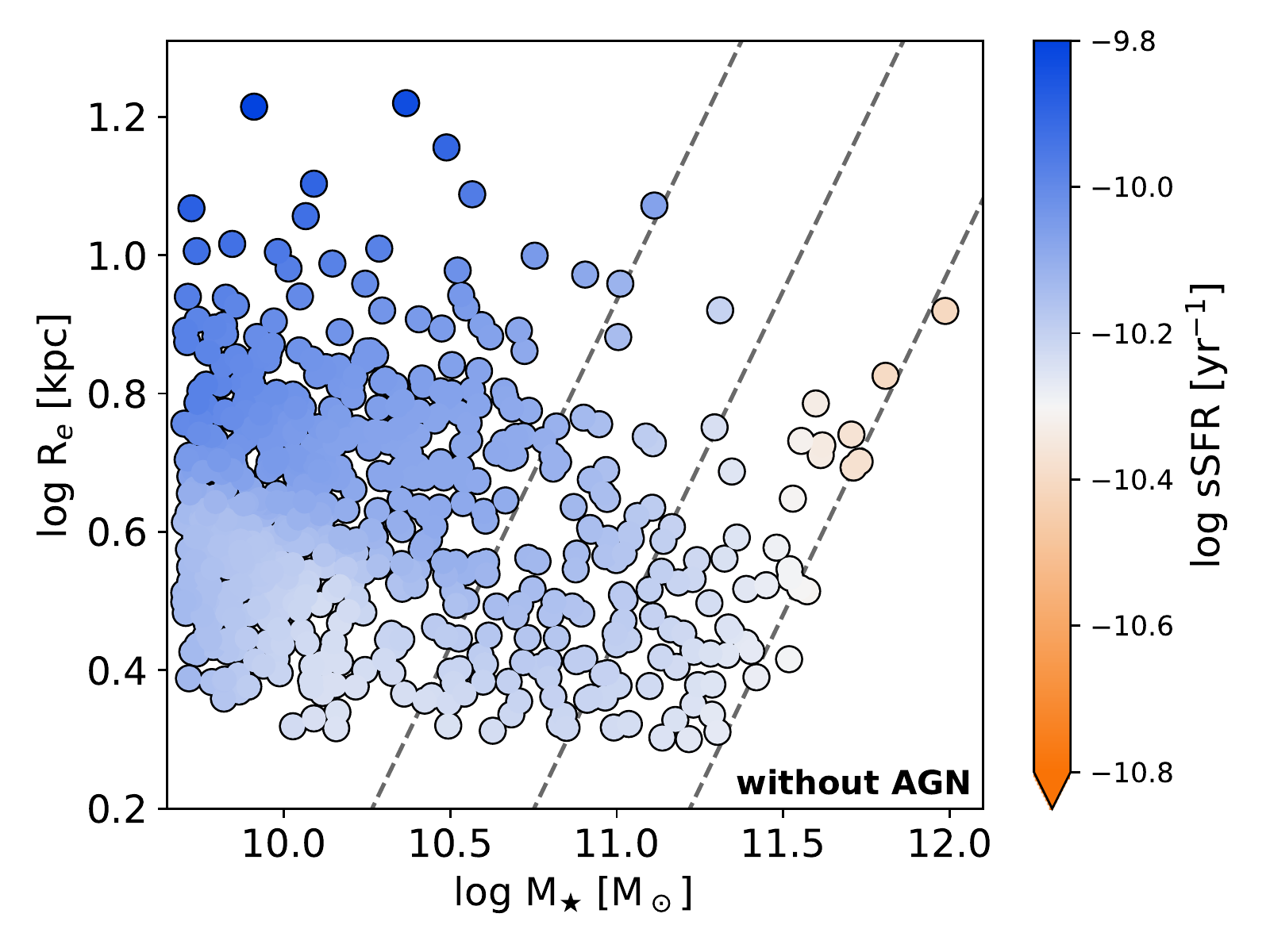}
     \end{center}
     \caption{Mass -- size plane in the Eagle cosmological simulation with (left panel) and without (right panel) AGN feedback. All galaxies more massive than 5$\times 10^9$\msun \ are shown here, color-coded by their sSFR (as measured over a 30 kpc aperture). As in Fig.~\ref{fig:msize}, the color-coding has been optimized with a LOESS smoothing scheme \citep{Cappellari13}. The color bar saturates at $\log$ sSFR $<-10.8$ to emphasize the two-dimensional color distribution. In the absence of AGN feedback, neither the morphology nor the color-coding follows the observational trends of Fig.~\ref{fig:msize}. Thus, as expected from a black hole-driven scenario, it is the effect of AGN feedback on the baryonic cycle of massive galaxies that determines their stellar population properties through a cumulative heating of the gas. For reference, dashed lines are lines of constant velocity dispersion assuming that galaxies are virialized. Note that the lines are not the same as in Fig.~\ref{fig:msize}, because the virial mass estimator described in \citet{Cappellari13} was calibrated with a fixed aperture of 1 \re, whereas the EAGLE measurements correspond to an aperture of 30 kpc.}
     \label{fig:simu}
\end{figure*}

In order to test this black hole heating scenario, we made use of the publicly available database of the EAGLE cosmological simulation \citep{EagleDB}. In particular, we compared their predictions for the mass--size plane in a simulated volume of 50 Mpc with and without black hole feedback. All measurements correspond to a fixed aperture of 30 kpc and we have only included galaxies with with a stellar masses M $_\star> 5 \cdot 10^9$ \msun. Although the EAGLE project does not offer predictions for the quenching time-scale as defined in Fig.~\ref{fig:msize}, it provides an estimation of the sSFR. In practice, both quantities are approximately equivalent since a longer quenching time-scale (bluer colors in Fig.~\ref{fig:msize}) implies a more extended star formation history and, therefore, a higher sSFR at z=0. The connection between quenching time-scale and sSFR is also empirically supported by the fact that, at fixed stellar mass or velocity dispersion, galaxies with lower sSFR \citep{Terrazas16} are also those exhibiting older ages and shorter quenching time-scales \citep{MN16b}. The comparison of the properties of the mass--size plane with and without the effect of AGN feedback is shown in Fig.~\ref{fig:simu}.

According to the EAGLE cosmological simulation, AGN feedback has two main effects on the mass--size plane. First, it helps to reproduce the size of the most massive galaxies that otherwise would be too small compared to what is observed. Moreover, the effect of AGN feedback is needed to efficiently quench star formation \citep{Crain15}. Otherwise, as shown in the right panel of Fig.~\ref{fig:simu}, massive galaxies continue forming stars at a much higher rate than observed \citep[e.g.][]{Scholtz18}.

Second, variations of galaxy properties along diagonal lines of constant velocity dispersion (dashed lines in Fig.~\ref{fig:simu}) emerge from this AGN feedback. In the absence of feedback, lines of constant sSFR are nearly horizontal (see Appendix~\ref{app:2}). Therefore, at least in this state-of-the-art cosmological simulation, the observed relation between stellar population properties and stellar velocity dispersion seems to be a consequence of black hole heating. 

It is important to note here that in Fig.~\ref{fig:simu} we selected all galaxies more massive than $5\times10^9$ \msun \ in the EAGLE simulation without applying any morphological selection, while the ATLAS3D sample shown in Fig.~\ref{fig:msize} only contains ETGs. Therefore, the comparison between these two figures is only an approximate assessment of the possible effect of black hole feedback in massive galaxies.

\subsection{Stellar population scaling relations: M$_\star$ vs $\sigma$}

In a $\Lambda$-CDM Universe, halo mass is thought to be the main quantity regulating galaxy evolution \citep[e.g.][]{Behroozi13}. It is still debated whether galaxy stellar mass or $\sigma$ is a better proxy for halo mass \citep[e.g.][]{Li13,Zahid16,Tinker17}, yet stellar population properties in massive ETGs correlate much better with galaxy velocity dispersion than with stellar mass \citep[e.g.][]{Trager00,Bernardi05,Gallazzi06,Graves09}. 

\citet{Barone18} have recently proposed that it is the gravitational potential $\log \Phi \propto \log \sigma$ that ultimately determines the stellar population properties, since galaxies with deeper potential wells, and therefore with higher escape velocities, are more efficient at retaining their stellar ejecta. However, as described above, it is unlikely that the gravitational potential plays a significant role in a mass range (M$_\star\gtrsim10^{10}$\msun) where the baryonic cycle is expected to be dominated by the effect of black hole feedback.  Moreover, X-ray gas temperatures (Fig.~\ref{fig:tgas}) and stellar population properties \citep{MN16} change with black hole mass at fixed $\sigma$, and therefore, at fixed $\Phi$ \citep[e.g.][]{Barone18}, assuming that galaxies are virialized systems. It is worth noting that, because of this virialized nature of galaxies, $\sigma$, $\phi$, and black hole mass are expected to track each other \citep[e.g.][]{vdb16}, although the details of these scaling relations depend on other processes such as galaxy mergers \citep[e.g.][]{Tapia14}, or black hole mass accretion history \citep[e.g.][]{simba}. It is precisely these second order effects over the dominant virial behavior of galaxies that we argue allows our relative assessment of the effect of black hole feedback.

Figure~\ref{fig:msize} presents a suggestive scenario to explain why stellar populations better correlate with $\sigma$ than with M$_\star$. At fixed stellar mass, galaxies exhibit a wide range of black hole masses and therefore of hot gas temperatures. Thus, if black holes are responsible for heating the gas above the virial temperature (set by the halo mass), and this change in the thermodynamics in dark matter halos affects the stellar population properties over time, a large scatter is expected. On the other hand, the stellar velocity dispersion does not only encode information about the halo mass, but also about the mass of the black hole, making $\sigma$ a suitable predictor for stellar population properties when direct black hole mass measurements are not available (see \S~\ref{sec:mss} for a discussion about the caveats of using $\sigma$ as black hole mass proxy.). Our joint analysis of direct black hole mass measurements and X-ray temperatures builds on previous observational efforts suggesting a connection between $\sigma$, black hole mass, and the star formation histories of massive galaxies \citep[e.g.][]{Wake12,Bell12}. This simple idea is able to unify in a coherent and physically motivated picture the observed properties of ETGs \citep[see e.g.][]{Cappellari16} and their dependence on galaxy mass \citep[e.g.][]{Terrazas16,Terrazas17} and velocity dispersion \citep[][]{MN16,MN18b}.

\subsection{Black holes and the circumgalactic medium}

Above, we have sketched a scenario where we have connected our observed correlation between the black hole mass (and, as a rough proxy, the velocity dispersion) and the X-ray emitting gas temperature in ETGs with the observed stellar population properties. This relation between black holes, stellar populations, and X-ray temperatures, although only probed close to the center of the halo ($r\sim$\re), may also extend out to large radii \citep[e.g.][]{{Fukazawa06,Diehl08}}. A remaining connection is needed to complete the story as to \textit{how} the black hole mass and stellar populations are coupled. We have mentioned that heating the gaseous halo, or CGM, can trap the metal-enriched stellar ejecta and prevent fresh gas from fueling further star formation. Because we have shown that energy injection from the black hole appears to have heated the gas in the central galactic regions, effects should be observable in the CGM and illuminate the quenching mechanisms further. 

Modern hydrodynamic simulations indeed show profound impacts on the CGM from black hole feedback. In the IllustrisTNG simulations, the amount of \textsc{Ovi} in the CGM is highly dependent on the $z\sim0$ black hole mass, as the metals are pushed further out into the halo and the halo is heated from accumulated black hole feedback \citep{Nelson:2018aa, Terrazas19}.  The EAGLE simulations also show profound effects on the CGM from black hole feedback; the gas mass fraction in the halo is reduced \citep{Davies:2019aa} in galaxies with more massive black holes. \citet{Oppenheimer:2019aa} predict that \textsc{Civ} is a good tracer of this CGM clearing, and galaxies at fixed stellar mass should exhibit a lower covering fraction of \textsc{Civ} in the halo with increased M$_\bullet$. Furthermore, \citet{Oppenheimer:2019aa} show that galaxies with lowered CGM gas mass fractions evolve secularly away from the ‘blue cloud’, as their galaxy samples were controlled to have histories free of major mergers over several Gyr.

Intriguingly, \citet{Burchett:2016aa} find a wide range of \textsc{Civ} column density in their highest mass galaxies, although no black hole information was known.  These tracers are best observed in ultraviolet (UV) absorption spectra of background quasars or galaxies and thus rely on fortuitous alignments of galaxies and UV-bright background source. \citet{Berg:2018aa} compared CGM properties of optically-selected AGN hosts versus a sample with no AGN and found no significant differences between the two populations.  However, the key parameter we focus on in this work and the parameter that drives the differences in CGM tracers within the simulations is the mass of the black hole, not its current accretion rate. Using velocity dispersion as a proxy for black hole mass could significantly aid building samples of quasar-galaxy pairs suitable for studying the impacts of black hole feedback on the CGM of massive ETGs.  

\section{Conclusions}~\label{sec:fin}

Massive ETGs stand as a benchmark for extra-galactic astronomy, challenging our understanding of galaxy formation even after decades of thorough study. Among all of the open questions, their quenching processes and baryon cycle in general are yet to be fully understood. We have presented here evidence of a relation between the temperature of the hot gas in massive ETGs and the mass of the central super-massive black hole. This result, in combination with the observed scaling relations between black hole and galaxy mass, stellar velocity dispersion, and stellar population properties, observationally supports the idea that black hole heating is a fundamental mechanism regulating the formation and evolution of massive galaxies. A comparison between the observed mass-size plane of ETGs and the predictions from the EAGLE suite of cosmological simulations also supports the key role of black hole heating in this simulation because black hole feedback is needed to reproduce realistic star formation histories of massive galaxies, ultimately leading to the observed scaling relations between galaxy mass, size, $\sigma$, and stellar population properties.

Our results roughly describe an interpretative framework where black hole feedback has a critical role in heating the gaseous halos around massive galaxies beyond the virial temperature, modifying the internal thermodynamics, and therefore affecting the cooling efficiency and the observed stellar population properties. In order to fully explore this scenario, it is necessary to increase the samples of objects with direct black hole mass measurements and to understand the connections to other important galaxy properties such as internal structure or large-scale environment. Future facilities such as {\it James Webb Space Telescope}, and in particular the upcoming 30-meter class telescopes, will be key in our understanding of the galaxy-black hole coevolution \citep[e.g.][]{astro2020}.

\section*{Acknowledgments}
We would like to thank the anonymous referee for their detailed and careful revision of this manuscript. IMN acknowledges support from the AYA2016-77237-C3-1-P grant from the Spanish Ministry of Economy and Competitiveness (MINECO) and from the Marie Sk\l odowska-Curie Individual {\it SPanD} Fellowship 702607. M.M. acknowledges support from the Spanish Juan de la Cierva program (IJCI-2015-23944).




\bibliographystyle{mnras}
\bibliography{bhot} 

\appendix
\section{The M$_\star$--$\sigma$ relation} \label{app:1}

\begin{figure}
     \begin{center}
     \includegraphics[width=8.7cm]{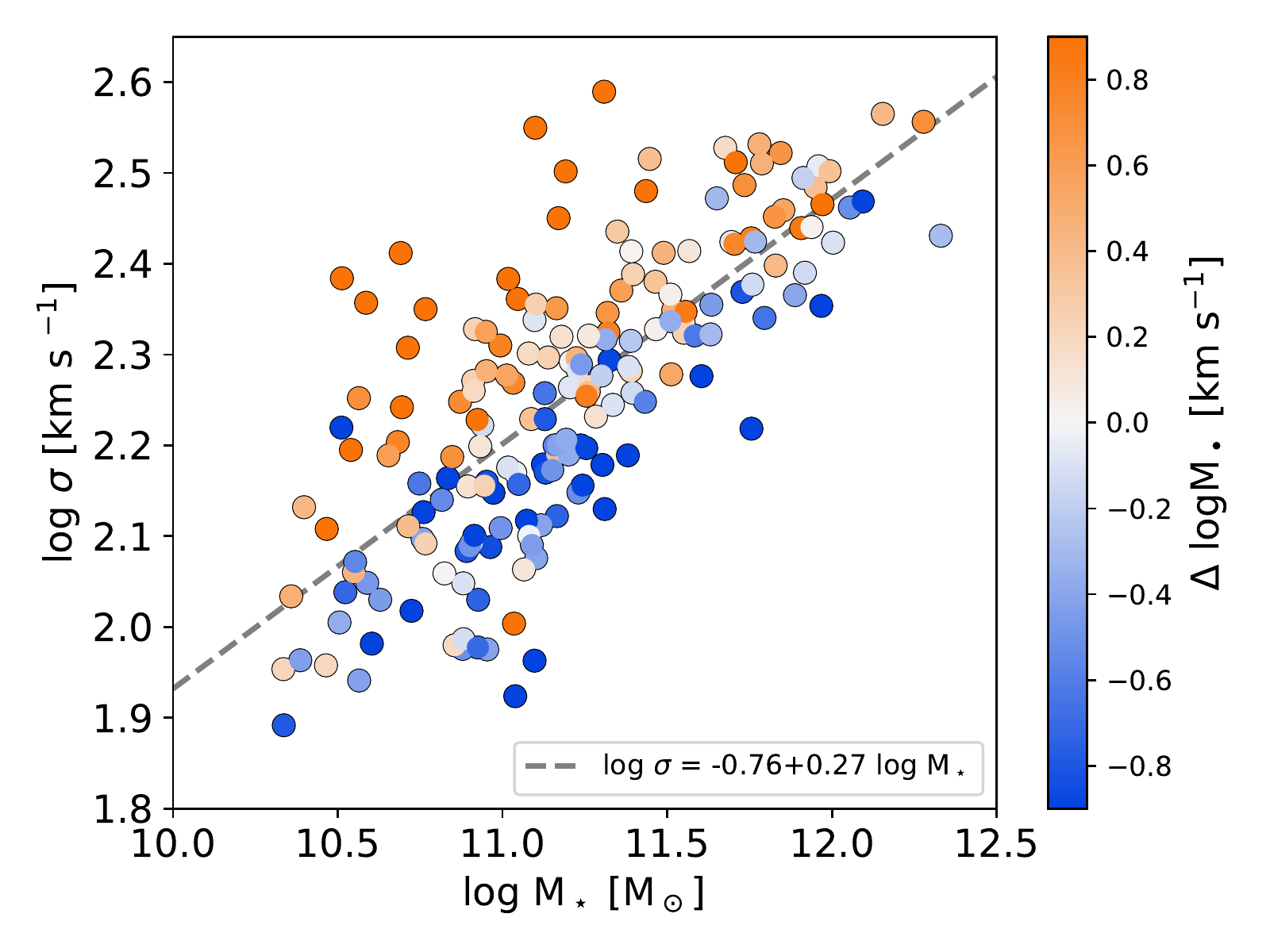}
     \end{center}
     \caption{ M$_\star$--$\sigma$ relation for the sample presented in \citet{vdb16}, where the best-fitting trend is shown as a dashed line. Symbols are color-coded as a function of their distance with respect to the average M$_\star$--\mbh \ relation. Dashed line indicates the average M$_\star$--$\sigma$ relation. This figure, along with Fig.~\ref{fig:scatter}, shows that for a given stellar mass galaxies hosting more massive black holes tend to also have higher stellar velocity dispersions.}
     \label{fig:app1}
\end{figure}

Fig.~\ref{fig:app1} shows the relation between the stellar velocity dispersion measured at 1\re \ and the total stellar mass, as measured by \citet{vdb16}. The color code is the same as in Fig.~\ref{fig:scatter} and indicates the distance with respect to the best-fitting relation. Galaxies that, at a given stellar mass, have higher stellar velocity dispersions also host more massive black holes. Therefore, the scatter in Fig.~\ref{fig:app1} can be used to probe, at first order (see caveats detailed in \S~\ref{sec:mss}), the possible effects of black hole feedback in the host galaxy when direct black hole mass measurements are not available.

\section{Mass--size plane in EAGLE without AGN feedback} \label{app:2}

\begin{figure}
     \begin{center}
     \includegraphics[width=8.7cm]{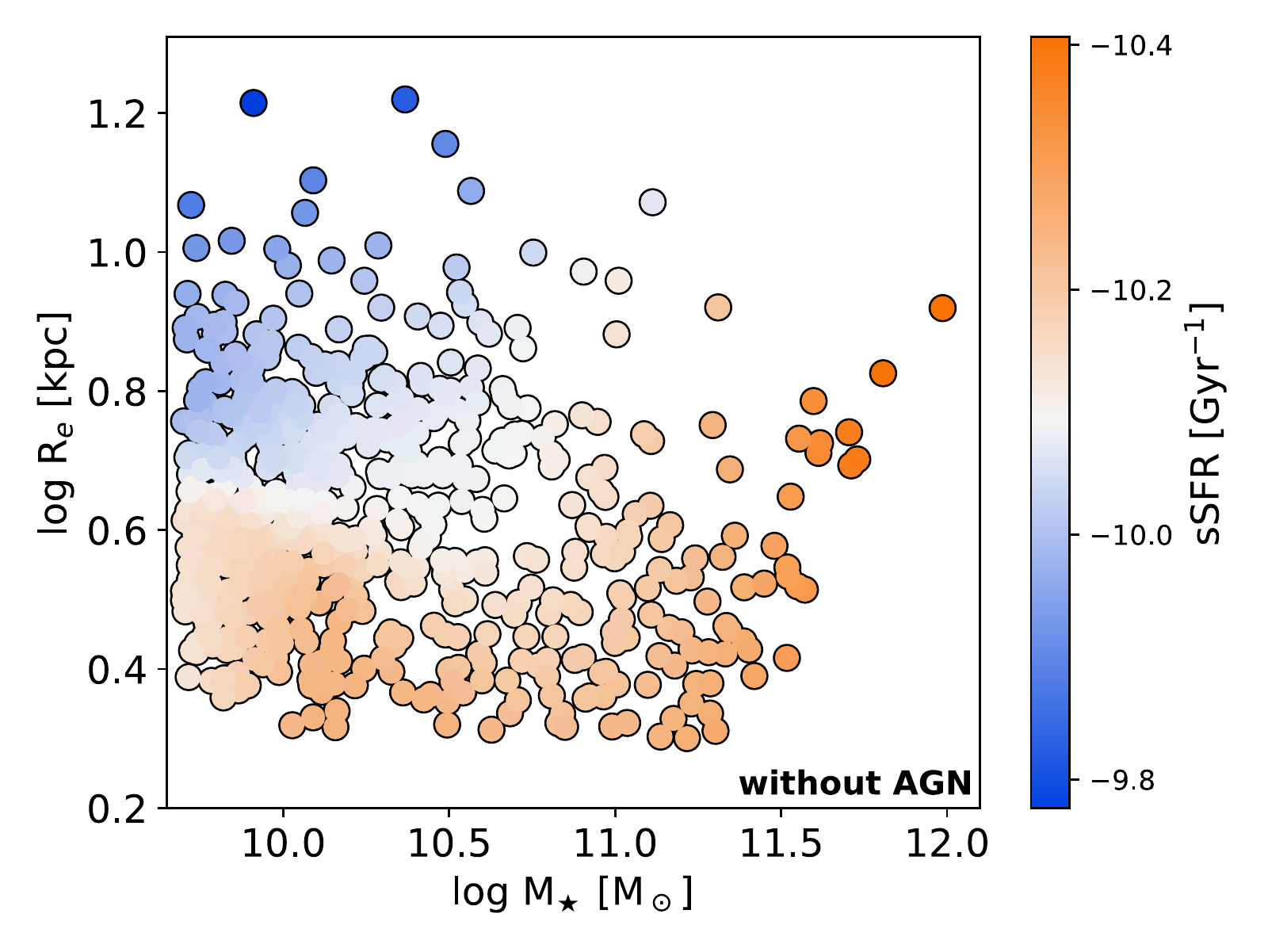}
     \end{center}
     \caption{Mass--size plane of the EAGLE simulation. This figure is the same as Fig.~\ref{fig:simu} but the color coding has been changed to show that, without black hole feedback, lines of constant sSFR are rather horizontal. This further supports the idea that black holes play a critical role in shaping the observed stellar population properties of massive galaxies, at least in the EAGLE cosmological simulation.}
     \label{fig:app2}
\end{figure}

Similarly to the right panel in Fig.~\ref{fig:simu}, Fig.~\ref{fig:app2} shows the mass--size plane of the EAGLE cosmological simulation without the effect of AGN feedback but in this case the color range has been adapted to show how the lines of constant sSFR are not diagonal, but roughly horizontal. This is due to the fact that, in EAGLE, sSFR is largely driven by the effect of black hole feedback. 

\bsp	
\label{lastpage}
\end{document}